 \documentclass[pmlr,twocolumn,10pt]{jmlr} 

\usepackage{booktabs}
\usepackage{siunitx}

\theorembodyfont{\upshape}
\theoremheaderfont{\scshape}
\theorempostheader{:}
\theoremsep{\newline}

\jmlrvolume{LEAVE UNSET}
\jmlryear{2023}
\jmlrsubmitted{LEAVE UNSET}
\jmlrpublished{LEAVE UNSET}
\jmlrworkshop{Machine Learning for Health (ML4H) 2023} 

\title[Assessing the Usability of GutGPT: A Simulation Study of an AI Clinical Decision Support System]{Assessing the Usability of GutGPT: A Simulation Study of an AI Clinical Decision Support System for Gastrointestinal Bleeding Risk}

\author{%
\Name{Colleen Chan}\Email{colleen.chan@yale.edu}\\
\addr Yale University, USA
\AND
\Name{Kisung You}\Email{kisung.you@baruch.cuny.edu}\\
\addr CUNY Baruch College, USA
\AND
\Name{Sunny Chung}\Email{sunny.chung@yale.edu}\\
\addr Yale School of Medicine, USA
\AND
\Name{Mauro Giuffrè}\Email{mauro.giuffre@yale.edu}\\
\addr Yale School of Medicine, USA
\AND
\Name{Theo Saarinen}\Email{theo\_s@berkeley.edu}\\
\addr University of California, Berkeley, USA
\AND
\Name{Niroop Rajashekar}\Email{niroop.rajashekar@yale.edu}\\
\addr Yale School of Medicine, USA
\AND
\Name{Yuan Pu}\Email{yuan.pu@yale.edu}\\
\addr Yale School of Medicine, USA
\AND
\Name{Yeo Eun Shin}\Email{yjshin@berkeley.edu}\\
\addr University of California, Berkeley, USA
\AND
\Name{Loren Laine}\Email{loren.laine@yale.edu}\\
\addr Yale School of Medicine, USA
\AND
\Name{Ambrose Wong}\Email{ambrose.wong@yale.edu}\\
\addr Yale School of Medicine, USA
\AND
\Name{Leigh Evans}\Email{leigh.evans@yale.edu}\\
\addr Yale School of Medicine, USA
\AND
\Name{Allen Hsiao}\Email{allen.hsiao@yale.edu}\\
\addr Yale School of Medicine, USA
\AND
\Name{Rene Kizilcec}\Email{kizilcec@cornell.edu}\\
\addr Cornell University, USA
\AND
\Name{Jasjeet Sekhon}\Email{jasjeet.sekhon@yale.edu}\\
\addr Yale University, USA
\AND
\Name{Dennis Shung}\Email{dennis.shung@yale.edu}\\
\addr Yale School of Medicine, USA
}

\begin{document}

\maketitle

\begin{abstract}
Applications of large language models (LLMs) like ChatGPT have potential to enhance clinical decision support through conversational interfaces. However, challenges of human-algorithmic interaction and clinician trust are poorly understood. GutGPT, a LLM for gastrointestinal (GI) bleeding risk prediction and management guidance, was deployed in clinical simulation scenarios alongside the electronic health record (EHR) with emergency medicine physicians, internal medicine physicians, and medical students to evaluate its effect on physician acceptance and trust in AI clinical decision support systems (AI-CDSS). GutGPT provides risk predictions from a validated machine learning model and evidence-based answers by querying extracted clinical guidelines. Participants were randomized to GutGPT and an interactive dashboard, or the interactive dashboard and a search engine. Surveys and educational assessments taken before and after measured technology acceptance and content mastery. Preliminary results showed mixed effects on acceptance after using GutGPT compared to the dashboard or search engine but appeared to improve content mastery based on simulation performance. Overall, this study demonstrates LLMs like GutGPT could enhance effective AI-CDSS if implemented optimally and paired with interactive interfaces.
\end{abstract}

\section{Introduction}
\label{sec:intro}

Large Language Models (LLMs), such as OpenAI's GPT-4, offer the next generation of foundational technology for clinical decision support using generative pretrained transformer architectures to provide a conversational interface for on-demand information retrieval and summarization. Its capacity to understand and respond to natural language queries have the potential to improve communication and enhance the efficiency of information retrieval, making it a valuable asset in everyday clinical practice. Notably, ChatGPT’s explanations of answers of USMLE sample questions showed high concordance and internal consistency among accurate explanations, highlighting its potential use as a didactic aid \citep{kung2023performance}. ChatGPT has been utilized to simulate conversations of breaking bad news by emergency medicine residents in an effort to prepare them for difficult conversations \citep{webb2023proof}. 

There is limited research on implementing artificial intelligence (AI) derived systems in clinical practice \citep{ethic}. Inadequate understanding of the human-algorithmic interaction, or in this case, clinician-LLM interaction, poses a major challenge to clinical implementation \citep{lee2021artificial}. The socio-technical challenge includes the issue of trust \citep{hengstler2016applied}, which for clinicians includes a need to understand AI systems' reasoning and a concern for legal liability \citep{lee2021artificial, kizilcec}. In fact, this need for an understanding of reasoning processes and transparency is not unique to the clinical field, but is fundamental in all AI applications \citep{glikson}.  Furthermore, suboptimal implementation may also lead to disruption in clinical workflows and inefficient use of clinician time, which is limited and expensive \citep{barrier}. Other structural issues also include difficulty capturing meaningful data, absence of adequate statistical expertise, and lack of training guidelines and opportunities \citep{lee2021artificial}. These challenges are amplified even further with the incorporation of new  technologies such as LLMs. 

Nevertheless, Epic Systems, the world's largest EHR vendor, announced earlier this year that it is partnering with Microsoft to integrate OpenAI's LLMs into its platform. The goal is to leverage AI to increase healthcare provider productivity through workflow automation and provide enhanced clinical decision support.  Pilot projects utilizing Microsoft’s OpenAI technology for automated message responses are already underway at a few major health systems including University of California San Diego Health, University of Washington Health, and Stanford Health Care.

Our multidisciplinary group validated a machine learning (ML) model predicting risk on the EHRs of patients presenting with acute GI bleeding (GIB), the most common cause of hospitalization for GI disorders \citep{shung2020validation}.
Risk models are clinically important, as practice guidelines recommend their use \citep{laine2021acg}.  To integrate the model into the EHR as part of the clinical workflow, we developed an interactive dashboard allowing clinicians to modify hypothetical patient covariates, such as lab values and medical history, and observe real-time changes in the predicted risk based on our model trained on local patient data in the Yale New Haven Health system. We also developed GutGPT, an AI chatbot interfacing with our validated ML risk prediction model and incorporating knowledge extracted from the latest clinical practice guidelines. For risk assessment, GutGPT provides the predicted risk via an integrated dashboard. For clinical management questions, GutGPT generates answers with evidence-based recommendations for patients with acute upper GIB.

To understand physician attitudes after exposure to our dashboard and/or GutGPT, we conducted our study in a simulation center using surveys measuring trust, acceptability, intention to use, and usability. A simulation center provides standardized conditions across different participants to enable fine-tuned adjustment of variables, allowing clearer assessment of measured outcomes. Medical simulation has proven to be an effective component of medical education, offering a controlled environment for trainees to become familiar with new technologies and learn about managing specific conditions \citep{ilgen2013technology}. Simulation environments are also valuable for studying new technologies that may pose risks if directly introduced into live clinical workflows \citep{rosen2008history}. 

Our study aims to assess GutGPT's efficacy in GIB decision support by measuring clinicians' trust in and acceptance of AI-CDSS for risk assessment. Additionally, we separately measure GutGPT's effect on mastery of clinical management knowledge.  Our goal is to better understand LLMs' impact on clinician-AI interaction through these endpoints.

\section{Methodology}
\label{sec:methods}

Both GutGPT and the interactive dashboard rely on a validated ML algorithm trained on an existing clinical dataset to predict GIB risk; see Appendix \ref{apd:mlmodel} for details. Through GutGPT's natural language interface, participants can ask questions about medical guidelines or the model’s predicted risk. Through the interactive dashboard, users can adjust patient covariates to predict risk of GIB for a hypothetical patient but natural language interaction is not possible. Details of GutGPT and the interactive dashboard are provided in Appendix \ref{apd:GutGPT} and Appendix \ref{apd:dashboard}, respectively.  

\begin{figure}[htbp]
 \floatconts
  {fig:flowchart}
  {\caption{
  Flowchart depicting the study protocol, where participants complete two phases. ``Interface'' refers to the interactive dashboard, and ``search'' refers to a general internet search engine. Participants complete surveys measuring outcomes before and after each phase. 
  }}
  {\includegraphics[width=1\linewidth]{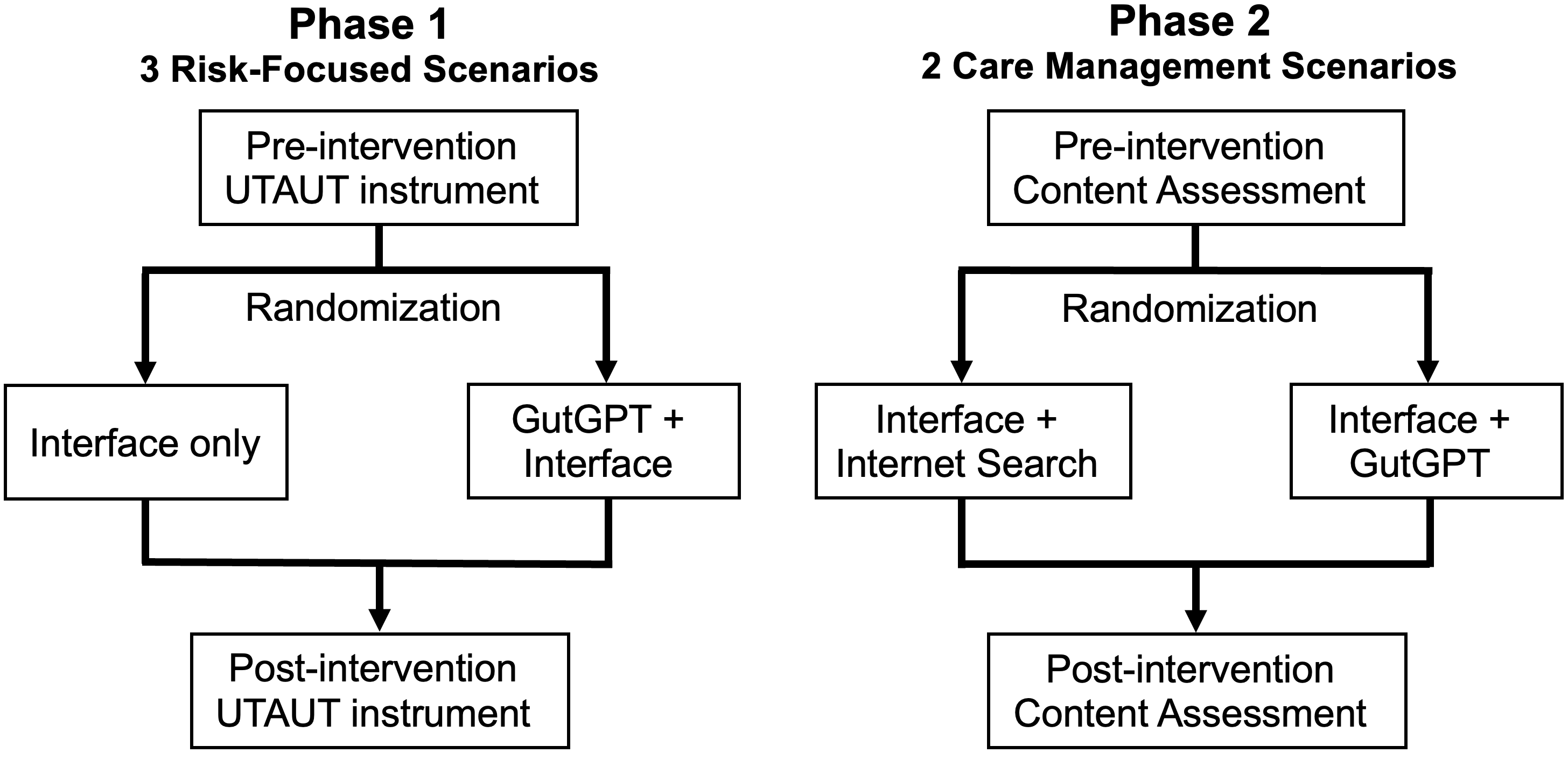}}
\end{figure}

We measured trust and acceptability using a survey adapted from two established instruments, the Unified Theory of Acceptance and Use of Technology (UTAUT) and the System Usability Survey (SUS), which have been previously applied in various industries like automated vehicles \citep{venkatesh2003user}. 

Beyond the  UTAUT established elements, our survey instrument also examined metrics on participants' trust in the system and their perceptions of its benefits, risks, and intelligibility. Additionally, the survey asked participants to rate their emotional response and overall attitude towards using the system in clinical practice. This survey was validated by physicians at the Yale School of Medicine to ensure the system's internal consistency for evaluating AI-CDSS \citep{ddwposter}. 

Emergency medicine and internal medicine physicians and medical students were enrolled and organized into small teams of two to four. Each team was provided with scenarios involving a high-fidelity Laerdal simulation mannequin on a gurney in a simulation laboratory mimicking a hospital examination room. The mannequin, equipped with a built-in microphone, displayed vital parameters, such as blood pressure and pulse. A computer terminal adjacent to the mannequin displayed a playground version of the Epic EHR populated with simulated patient data, including past medical history and medications; see Appendix \ref{apd:epic} for details.

The study has two phases: the first on evaluating GutGPT's effect on trust and acceptability, and the second on its effect on knowledge of clinical management. \figureref{fig:flowchart} illustrates our study protocol. 

In phase 1, participants complete a pre-simulation survey measuring their trust levels in AI-CDSS. After an AI-CDSS educational module, teams are randomized to GutGPT with the interactive dashboard or the interactive dashboard alone. They complete three ``Risk'' scenarios, where they must assess patient risk and decide if the patient should be discharged, admitted for observation, or admitted for in-hospital management. Post-simulation trust surveys are then administered.

In phase 2, participants complete an online pre-educational assessment testing management content from GIB management guidelines. They are re-randomized to GutGPT with the interactive dashboard or the interactive dashboard and online resources, such as internet searches and traditional clinical information sites. They then navigate two ``Content'' scenarios managing GIB cases, where they are tasked with making decisions regarding initial care management for patients with acute upper GIB situations. A post educational assessment about GIB management is then administered; the assessment can be found in the Supplementary Materials. 

After completing both phases, participants are debriefed on their experience. Surveys and educational assessments taken before and after are compared. Screen and video recordings of GutGPT use are captured and analyzed alongside qualitative feedback from the debrief to evaluate the interface's usability.

The study has been evaluated and deemed exempt by the Institutional Review Board at our institution.  

\section{Preliminary Results}
\label{sec:results}

\begin{figure}[htbp]
\floatconts
  {fig:alpha}
  {\caption{Measurement of reliability for adapted UTAUT metrics.}}
  {\includegraphics[width=1\linewidth]{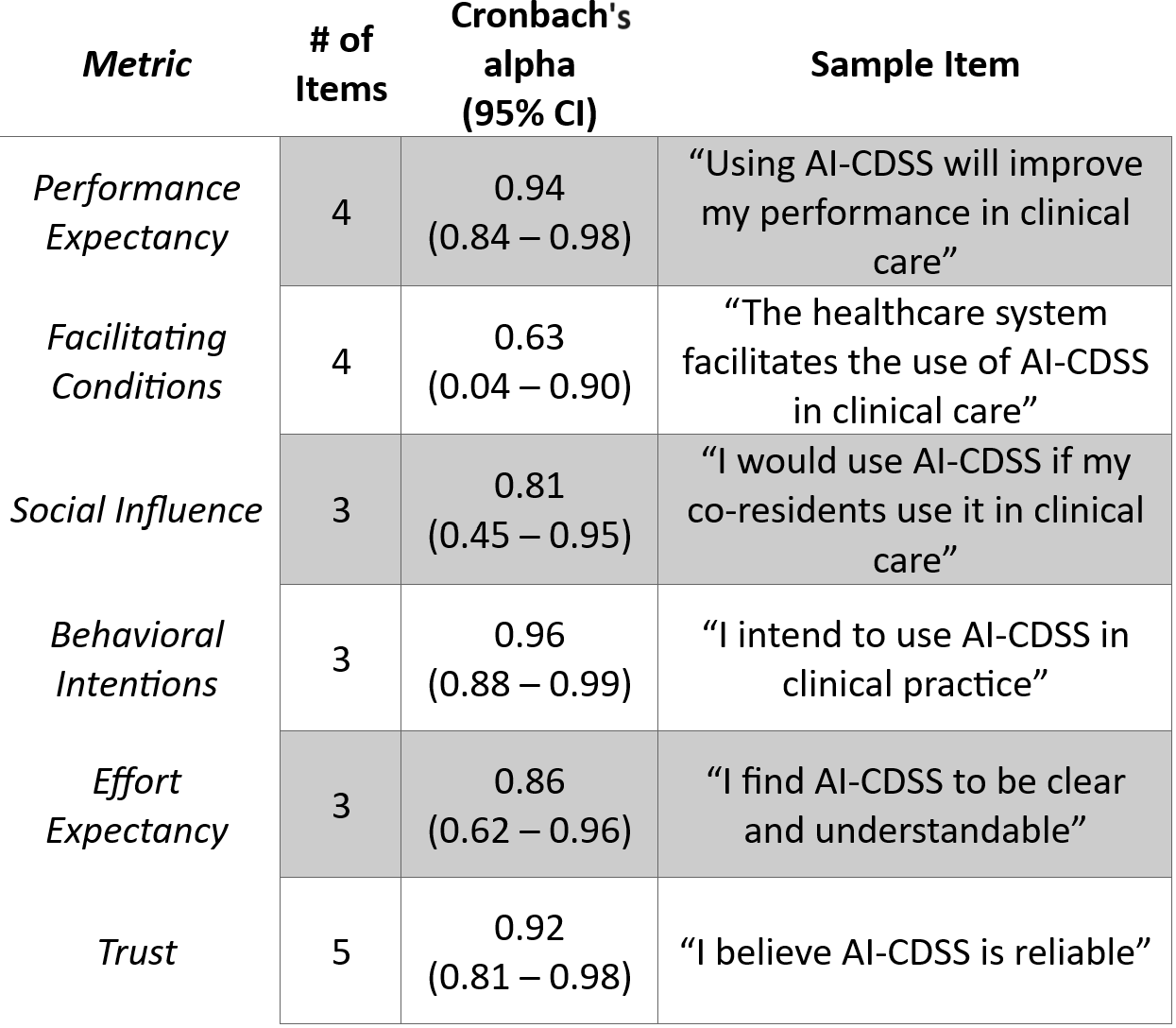}}
\end{figure}

\begin{figure}[htbp]
\floatconts
    {fig:metric}
    {\caption{Adapted UTAUT metrics for each arm (GutGPT vs dashboard) before (``Pre'') and after (``Post'') simulation. Higher Likert scale values represent more positive perceptions. Error bars represent $\pm 1$ standard deviation. 
    }}
    {\includegraphics[width=\linewidth]{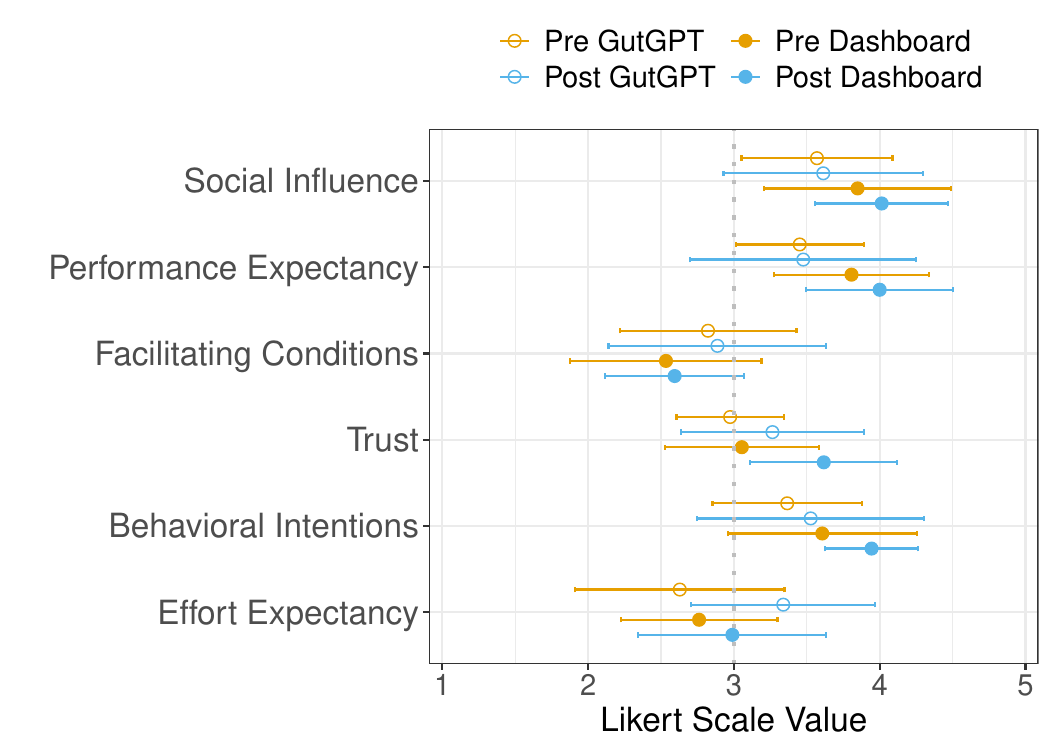}}
\end{figure}

Our study has enrolled 55 participants so far. In  phase 1, 31 were randomized to the dashboard arm and 24 to the GutGPT arm. In phase 2, 23 were randomized to the internet search arm and 28 to the GutGPT arm. The study remains ongoing and continues to actively enroll participants. Preliminary results are presented below.

We utilized the survey described in the previous section to examine trust and acceptability. The survey instrument's Cronbach's alpha reliability was re-validated \citep{tavakol2011making}.  \figureref{fig:alpha} shows that almost all Cronbach's alphas are greater than 0.8, suggesting high internal reliability.  The full survey can be found in the  Supplementary Materials. 

\figureref{fig:metric} shows UTAUT metric trends.  In general, after exposure to the simulation, participants in both arms increased their intention to use AI-CDSS, particularly for the dashboard arm.  Trust also increased for both groups.  Interestingly, Effort Expectancy, which corresponds to perceived ease of use, particularly increased for GutGPT arm participants.  Given the small sample size and ongoing recruitment, no statistical testing was conducted.

On content mastery, participants in both the Dashboard and GutGPT arms generally showed improvement (see \figureref{fig:education}).  

Dashboard arm participants viewed the system as a clinical assistant, while GutGPT participants perceived the system as helpful with patient triage. For both arms, the main concern was the perception that AI systems did not consider social, emotional, and physical nuances that contribute to clinical decision-making. 

 \begin{figure}[htbp]
\floatconts
    {fig:education}
    {\caption{Percentage of educational assessment questions answered correctly, averaged across participants by randomization group. Error bars represent $\pm 1$ standard deviation.}}
    {\includegraphics[width=1\linewidth]{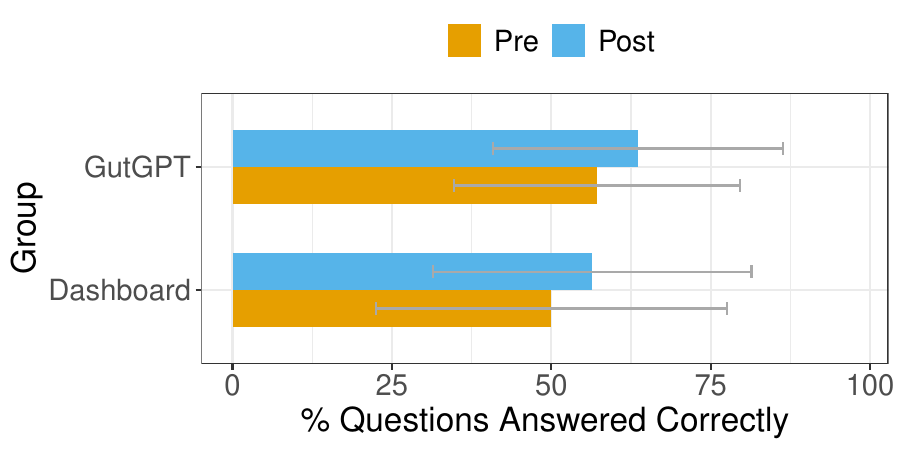}}
\end{figure}

\section{Discussion and Future Work }
\label{sec:disc}

This study demonstrates the use of medical simulation to evaluate an LLM-based chatbot for acute GIB management without compromising patient safety. 

Simulation enables controlled testing of new AI-CDSS before deployment and comprehensive evaluation of clinician attitudes towards its use. LLMs like GutGPT showcase the potential to test the use of AI in high-stakes clinical scenarios for clinical decision support. As these technologies rapidly progress, simulation studies are a helpful setting for understanding the safety risks and optimizing the user experience to promote responsible use and to maximize positive clinical impact. 

Preliminary results suggest GutGPT and the interactive dashboard increase knowledge acquisition and maintain postiive perceptions of trust in simulated scenarios. However, the impact on trust and acceptance is mixed.  Effort expectancy, which measures the ease of use, appears to increase with the GutGPT use. However, this may not necessarily translate to increased trust or intention to use.  

Limitations include potential bias from the sequential study design and limited generalizability of simulation. The sequential phases with separate randomizations could underestimate effects in the second phase due to increased familiarity with the LLM interface. However, we believe providing an overview of the systems before simulation minimizes this effect. In addition, the simulation scenarios were specifically designed to measure two distinct and separate aspects: the effect of GutGPT on clinician trust in its risk assessment and its impact on educational mastery regarding upper GIB management. While simulation differs from real-world practice, we believe it is an appropriate setting to evaluate new AI systems with unknown safety risks without disrupting existing care environments.

Future directions include performing a comparative analysis of different LLMs, identifying optimal temperature parameters to minimize hallucinations, and fine-tuning LLM architectures to better produce clinically relevant responses. Virtual/augmented reality could also improve accessibility, scalability, and customization of further simulation studies.

\acks{}

We are grateful to Tony Polcaro, Joy Grabow, and Jeff Hoffman of the Yale Center for Healthcare Simulation for their technical support of the ongoing trial. 

\bibliography{jmlr-sample}

\appendix

\section{Epic Electronic Health Record Playground}
\label{apd:epic}

The study utilizes a test version of the Epic health record, which creates an electronic environment incorporating simulated patient data used during the simulation.  The test version retains all the standard Epic capabilities during clinical practice, including visualization of the simulated patient's laboratory data, past medical history, medication history, and social history.   Since it operates within a test setting, the software prevents order entry and the addition of new data. \figureref{fig:epic} displays screenshots of this test version.

\begin{figure*}[htbp]

\floatconts
{fig:epic}
{\caption{Screenshot of the Snapshot screen (top) and Results screen (bottom) of the simulated patient on the playground version of Epic. The style and format is similar to the Epic version normally used  by clinicians in real life practice. }}

{
  \includegraphics[width=\linewidth]{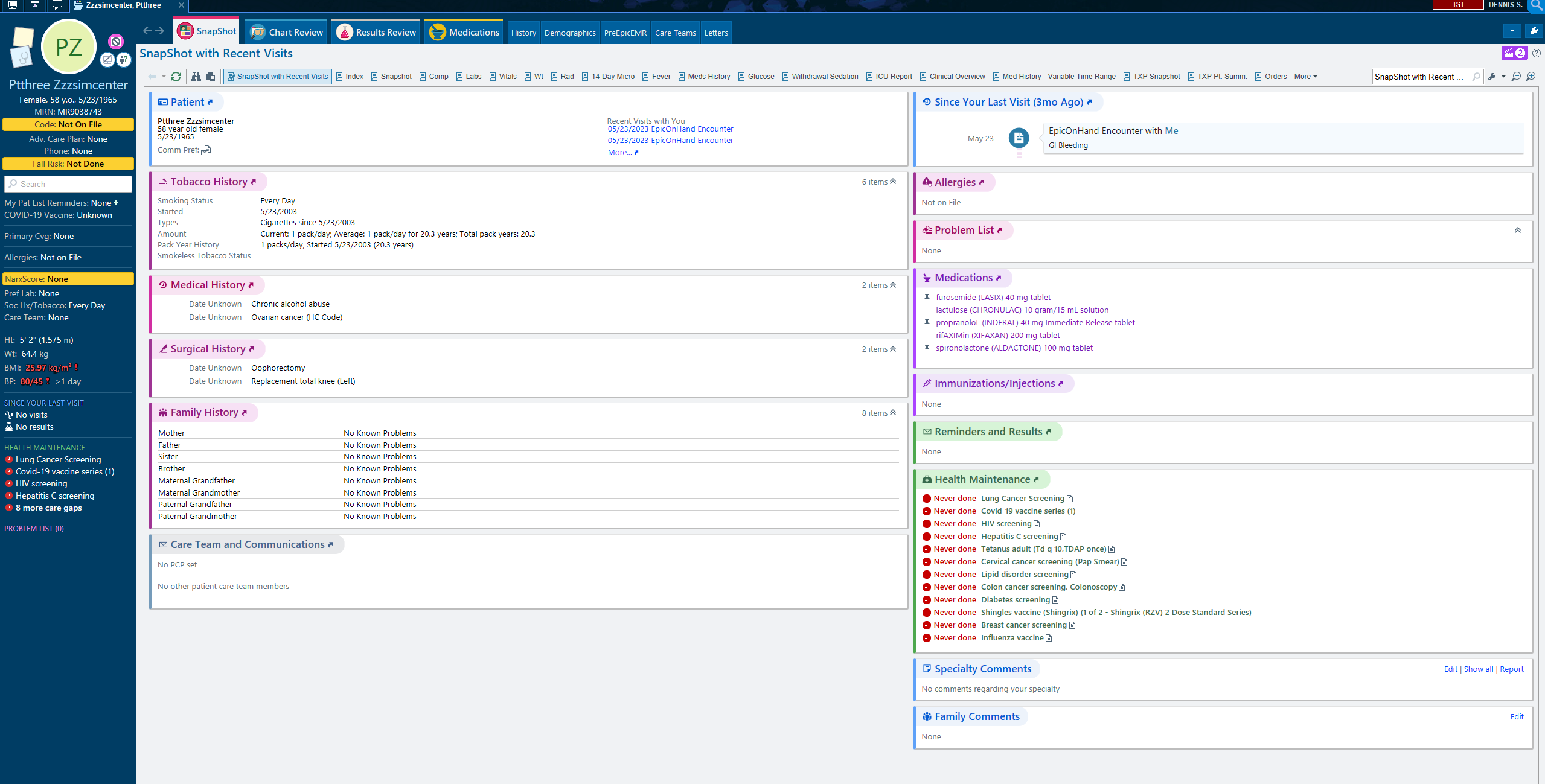} \\ 
  \vspace{1em} 
  \includegraphics[width=\linewidth]{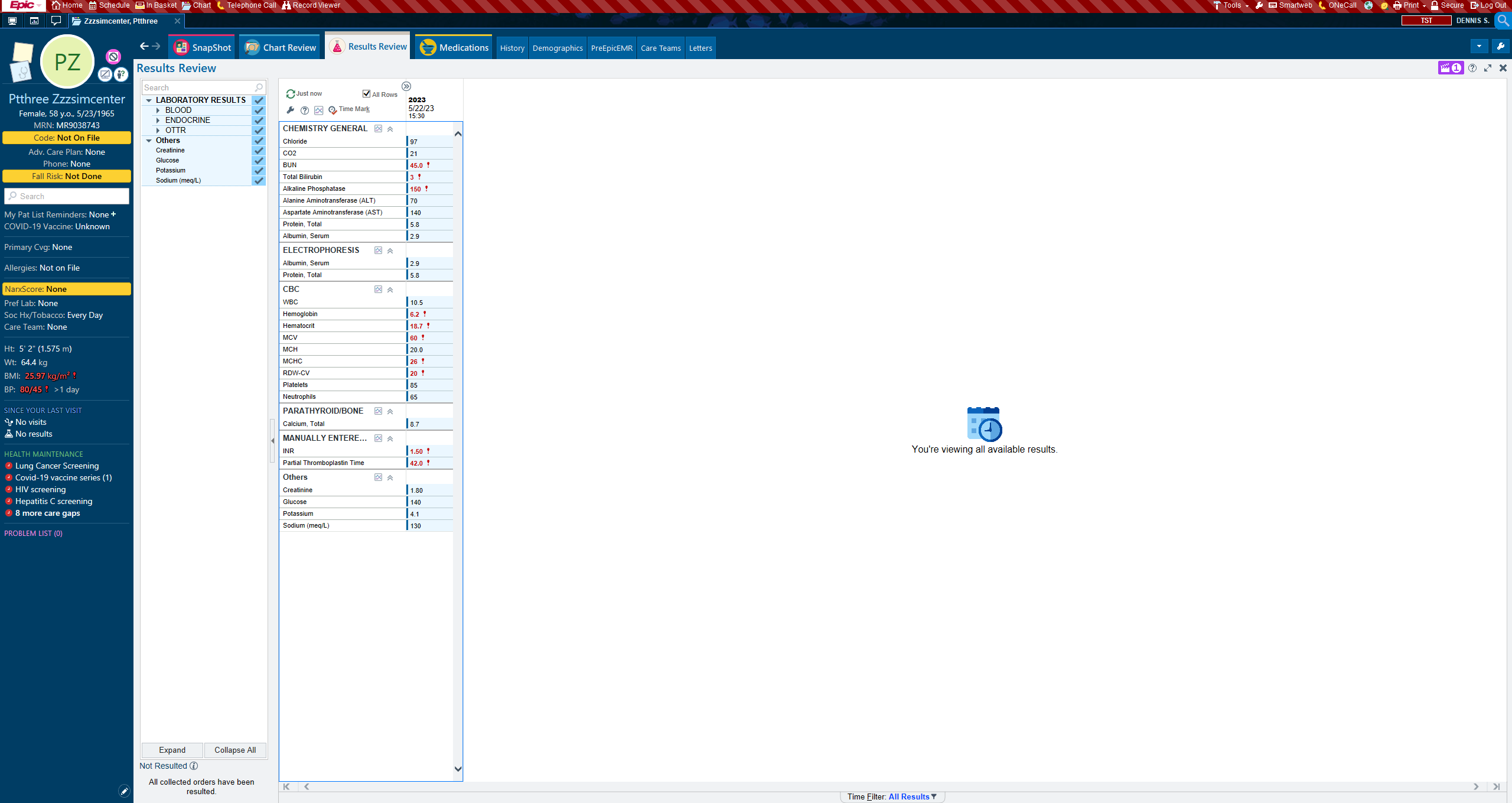}  
}

\end{figure*}

\section{Machine learning model}
\label{apd:mlmodel}

Both the interactive dashboard and GutGPT chat interface rely on an underlying model described below. 

The ML model was developed on the EHR data of patients in the Yale New Haven Health system presenting with reported or witnessed signs or symptoms of acute overt GIB. The input variables include demographics (age and sex), nursing assessment variables, lab test results, personal medical history, and medication classes in the form of Clinical-Classification-Software (CCS) codes. The model's outcome is a composite binary measure, encoded as 1 if a hospital-based intervention (red blood cell transfusion, endoscopic or hemostatic intervention) was required or if there was 30-day mortality and as 0 otherwise. Several ML and deep learning estimators, including random forests with honesty \citep{wager2018estimation}, gradient boosted trees \citep{chen2016xgboost}, supervised 2-layer and 5-layer neural networks \citep{rumelhart1986learning}, LASSO regression \citep{tibshirani1996regression}, and embedding methods, including principal components analysis \citep{pearson1901liii}, canonical correlation analysis \citep{hotelling1992relations}, variational autoencoders \citep{kingma2013auto}, and 2-layer neural networks, were explored. 

The selected model first applied separate LASSO regressions on the patient’s medical history and medication classes to reduce the dimensionality of the data. Random forests with honesty were subsequently applied to the variables yielding non-zero coefficients, in addition to the demographics, nursing assessment, and lab test variables. This model exhibited the highest true negative rate at a true positive rate of 99\% and an AUC exceeding 0.9.

\section{Details of GutGPT}
\label{apd:GutGPT}

 \begin{figure*}[htbp]
\floatconts
    {fig:gutgpt}
    {\caption{    Screenshot of the workstations of two simulation participants randomized to the GutGPT group.
    }}
    {
    \includegraphics[width=1\linewidth]{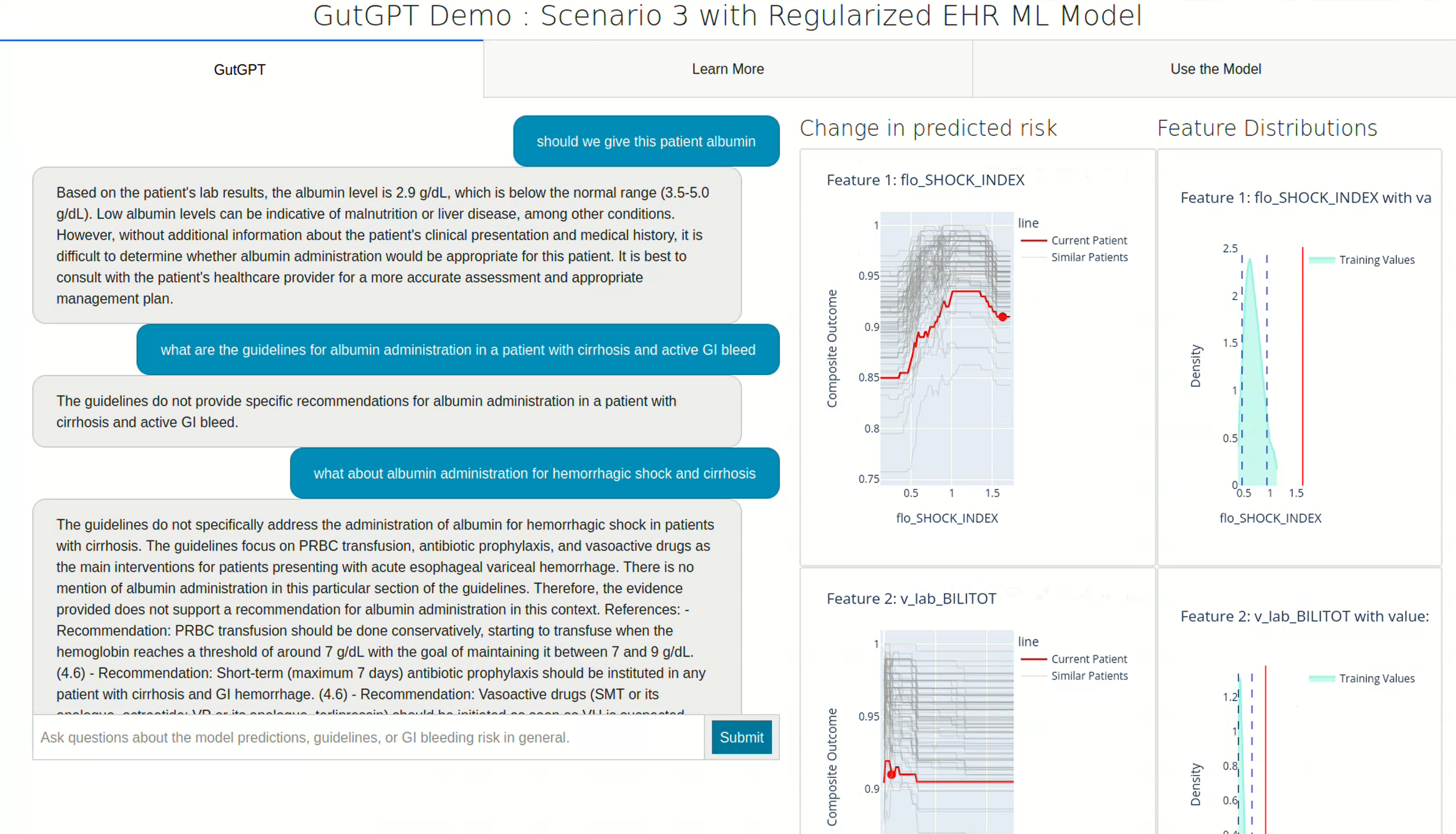} \\
    \vspace{0.5cm}
    \includegraphics[width=1\linewidth]{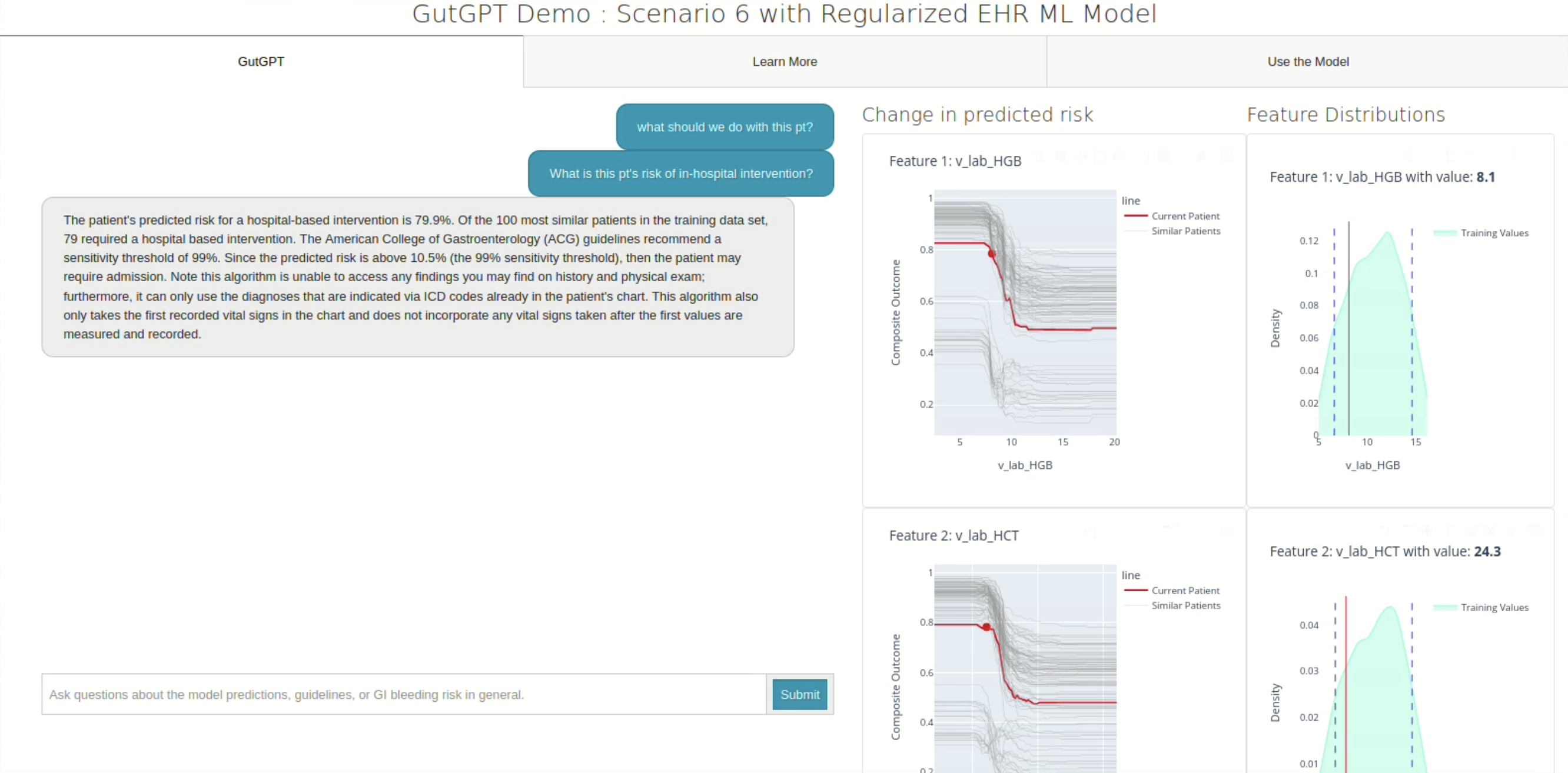}
    }
\end{figure*}

\begin{figure*}
    \centering
    \includegraphics[width=1\linewidth]{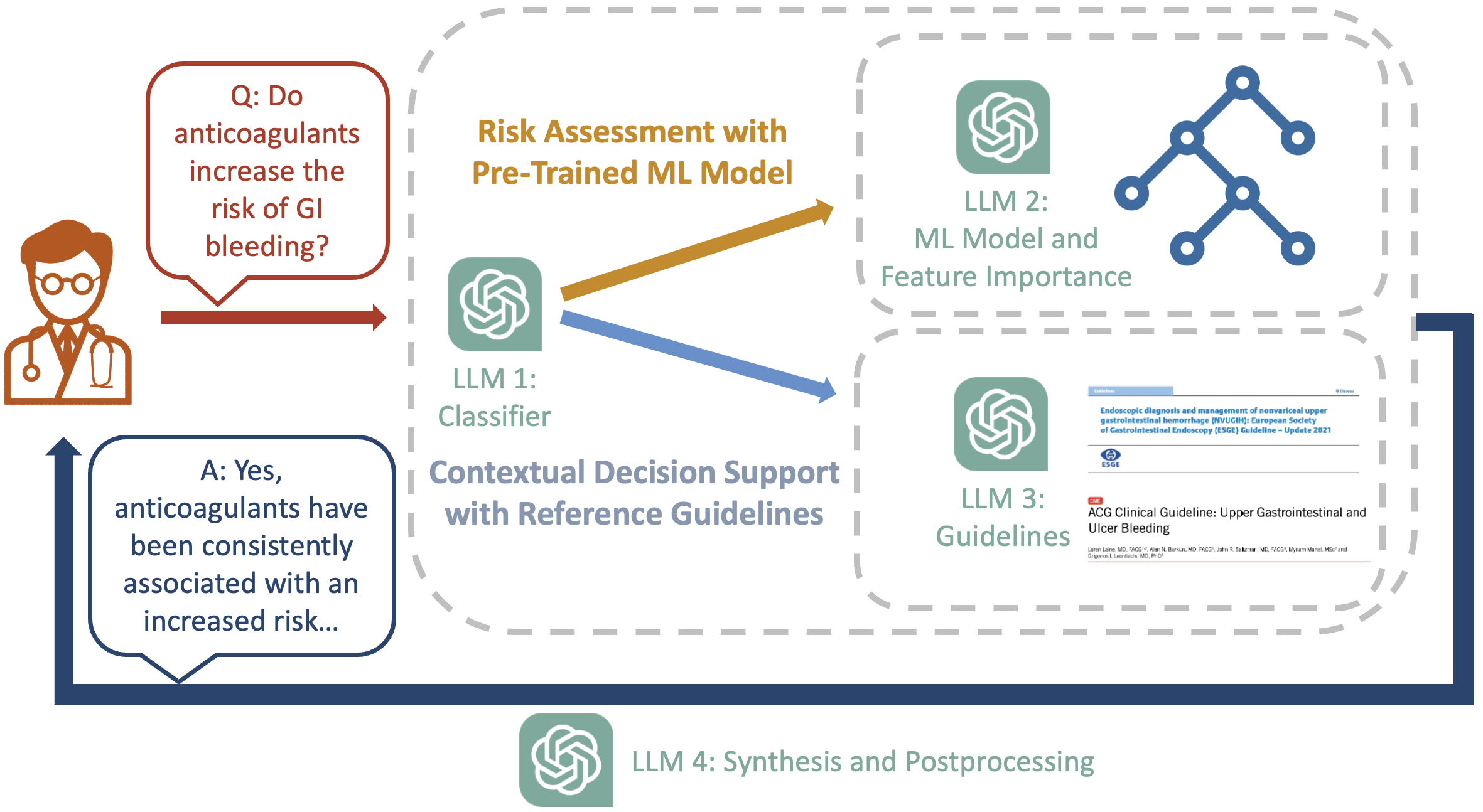}
    \caption{GutGPT query workflow}
    \label{fig:gutgpt_workflow}
\end{figure*}

GutGPT utilizes OpenAI's GPT-3.5 Turbo 16k model API to respond to user queries using in-context learning. When a question is typed, it undergoes a multi-step process (Figure \ref{fig:gutgpt_workflow}). 

First, a classifier LLM categorizes the query into one of three categories below, each of which uses a separate model. Several examples are provided to the classifier LLM as context. 

\begin{enumerate}
\item  \textbf{Model LLM}: If the query pertains to the predicted risk of GIB or important features contributing to the prediction, the prompt is directed to the ``Model'' LLM. This LLM retrieves the predicted risk from the ML model described in Appendix \ref{apd:mlmodel}. 
\item \textbf{Guidelines LLM}: If the query concerns medical guidelines, the prompt is directed to the ``Guidelines'' LLM, which retrieves the most relevant excerpt from a comprehensive GIB guidelines document to provide context and answers the query with relevant citations. 
\item \textbf{General LLM}: For questions unrelated to GIB or general GI queries, the prompt is directed to the ``general'' LLM, which is provided with only context that it has to answer GI-related questions for medical professionals.
\end{enumerate}

The patient's EHR data is automatically loaded at launch, serving as context for all queries except those directed to the classifier. For queries spanning multiple categories, a final ``synthesizer'' LLM generates the response. In all other instances, the response is directly outputted. \figureref{fig:gutgpt} displays examples of the GutGPT chat interface during the simulation scenarios. The model LLM and guidelines LLM are described in further detail below. 

The model LLM retrieves the predicted risk of a hospital-based intervention from the underlying ML model described in Appendix \ref{apd:mlmodel}. If the risk falls below the 99\% sensitivity threshold, the prediction is considered ``very low risk'' by the American College of Gastroenterology and ``not very low risk'' otherwise.
The model also provides the three most significant features contributing to the prediction if asked.

The guidelines LLM has access to a comprehensive text sourced from the guidelines of the American College of Gastroenterology for the management of upper GIB \citep{laine2021acg}. The guidelines are formatted into sections labeled pre-endoscopic management, endoscopic management, summary of evidence, recommendations, and conclusions. For pre-processing, the sections are segmented and transformed into vector embeddings using OpenAI’s text embedding model. These vector embeddings are subsequently saved as a highly optimized database called vector stores. When a clinician types a query, it is also converted into a vector embedding using the same text embedding model. Then, a similarity search of the query's vector embedding is performed between the query vector embedding and those in the database to retrieve the portion of the guidelines text most relevant to the query. This retrieved text serves as context for the prompt supplied to the LLM model along with formatting guidance and the patient's EHR data. The original query and this engineered prompt are finally fed into the GPT-3.5 Turbo 16k model to generate a response for the user.

\section{Details of Interactive Dashboard}
\label{apd:dashboard}

\begin{figure*}[htbp]
\floatconts
    {fig:dashboard}
    {\caption{
    Screenshot of the workstation of a simulation participant randomized to the interactive dashboard group.
    }}
    {
    \includegraphics[width=1\linewidth]{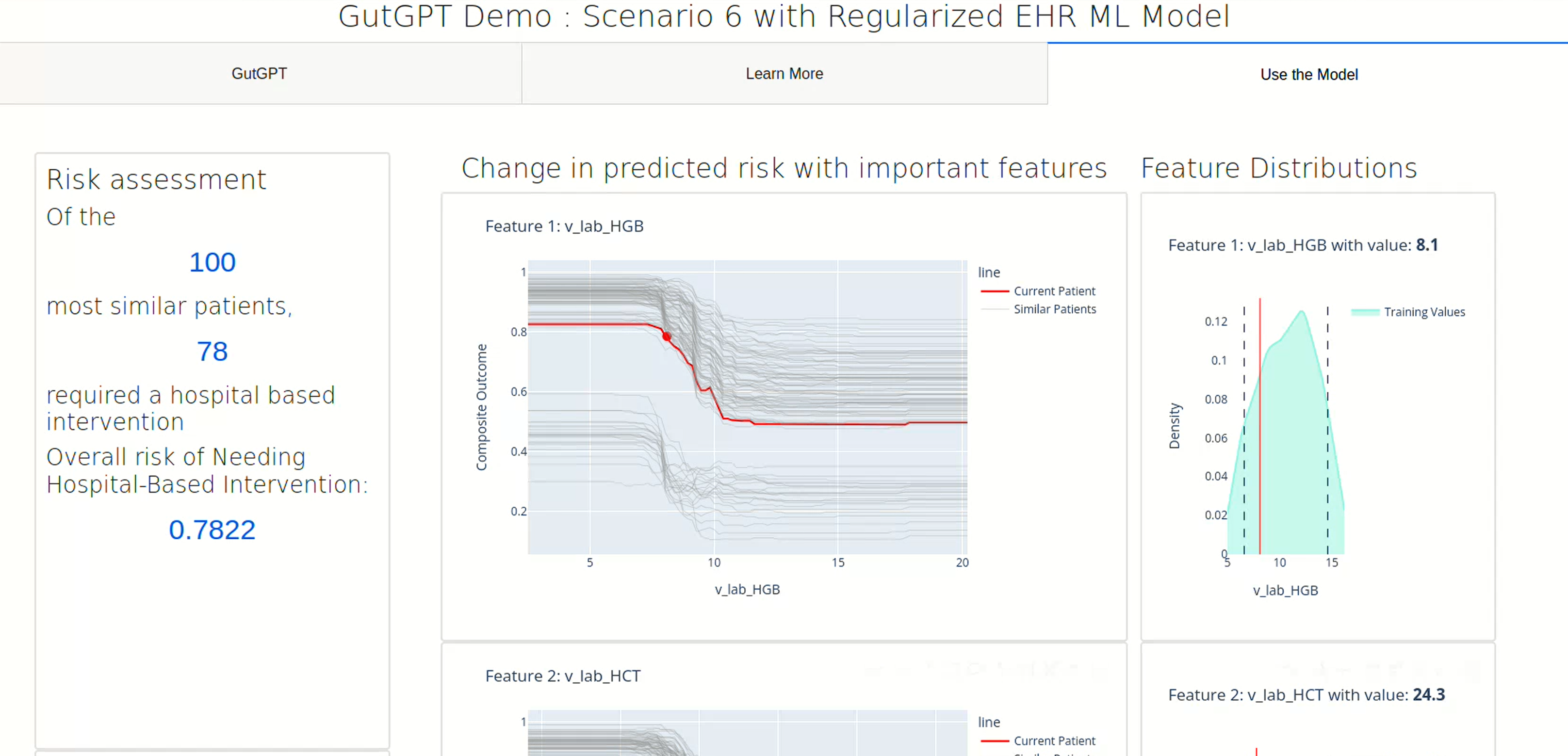}
    \includegraphics[width=1\linewidth]{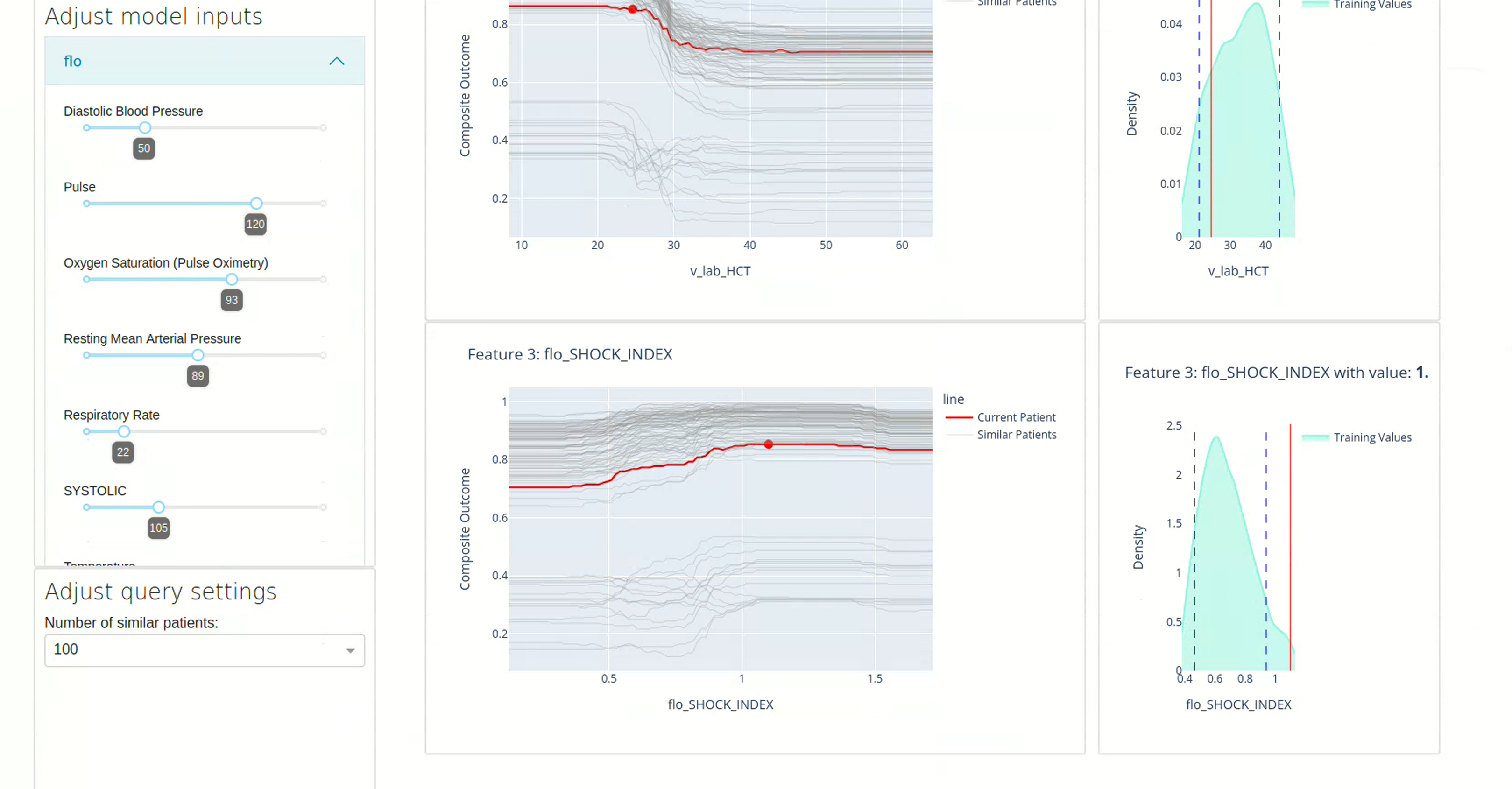}
    }
\end{figure*}

The interactive dashboard consists of two tabs: ``Learn More'' and ``Use the Model'' (see \figureref{fig:dashboard}).

The ``Learn More'' tab displays interpretability plots for the ML model used by GutGPT. Users can select any model covariate (e.g., demographics, lab values, medications) and view univariate or bivariate partial dependency plots (PDPs), individual conditional expectation (ICE) plots, and accumulated local effects (ALE) plots. These plots show how the selected covariates affect the model's predicted risk \citep{molnar2020interpretable}. PDPs show the marginal effect of one or two features on a model's prediction.  ICEs show how predicted risks change as a function of a feature for each individual observation, allowing one to observe heterogeneity among prediction paths.  ALE plots shows the expected change in the predicted response as a single feature value is varied over its range while averaging out its interaction with other features, making it less sensitive to correlated predictors than PDPs.  Incorporating interpretability plots allows users to better understand the ML model's decision-making process, ensuring it aligns with their clinical mental model. 

In the ``Use the Model'' tab, users can modify patient covariate values and observe in real-time how the predicted risk of a hospital-based intervention changes. In addition, an ICE plot for the hypothetical patient is displayed for the patient's top three most important features alongside the 100 patients from the training data most similar to them. Patient similarity is determined by the proportion of trees in the random forest model that share the same leaf node as the hypothetical patient. 
Feature importance is determined by ranking the inflection (rate of change) in predicted risk of the PDP function for each feature at the hypothetical patient's value. Histograms superimposed with the hypothetical patient's value for each of the three important features are also shown adjacent to the ICE plots. 

\end{document}